# An easily prepared, simple information storage and display device based on triboelectric enhanced mechanoluminescence phenomenon and droplet-luminescence

*Changhui Song, Shicai Zhu, Liran Ma\*, Yu Tian, Jianbin Luo\**

Changhui Song, Shicai Zhu, Dr. Ma., Prof Tian., Prof. Luo.
Shuangqinglu 30, Tsinghua University, 100084, China
E-mail: maliran@tsinghua.edu.cn, luojb@tsinghua.edu.cn

One easy-make, flexible information storage and display device is prepared. The device show memory effect, which can record the friction trace in real time through the triboelectric effect, and then reproduce the trajectory under the excitation of any kind of polar liquid droplet, completely display the triboelectric area. The explanation mechanism of this phenomenon is proposed, and the main influencing factors of the frictional electrification between solid materials and flexible substrates are determined by systematic experiements. This phenomenon connects mechanoluminescence, electroluminescence, and triboelectrification. It is of great significance to the research of triboelectricity and mechanoluminescence, and is expected to play an important role in display, information storage and encryption.

## 1. Introduction

Mechanoluminescence material was discovered as early as 1605[1]. It is a functional material that can directly convert mechanical energy into light and can be excited under the action of various forms of mechanical energy[2], such as friction, pressure, stretching, shearing, etc. In the early days, researchers believed that the luminescence of materials must be accompanied by different degrees of damage to the materials themselves, so it did not attract enough attention. However, around 1999, CN Xu et al. discovered that doped ZnS[3] materials can emit light in the elastic strain range, which greatly attracted the attention of researchers, since strain in the elastic range means the material can be reused without damage[4]. More important, unlike other Mechanoluminescent materials, the doped ZnS does not require any pre-irradiation and can maintain the luminous intensity even after one hundred thousand repeated cycles[5]. Therefore, the research on this material is carried out systematically. Chandra et al. studied on the mechanism and proposed a pressure-excited carrier detrapping model[6] to explain the reason for the luminescence of doped ZnS, and named it as elastico-mechanoluminescence. Jeong and others have made great contributions to the application. They not only achieved color manipulation[7], but also demonstrated the huge application potential of doped ZnS materials in display[8] and anti-counterfeiting.

Among the numerous applications of doped ZnS materials, flexible array sensors that can be used to display handwriting[9, 10] grab the most attentions. Researchers use it as a tool to convert force into the light. The signer's stroke strength and signing speed can be converted into light intensity in real-time[9], then recorded by the CCD camera on the back of the material, and then infer the signature's handwriting speed and strength. However, this transposition requires stricter control of the material's thickness because writing and light intensity recording is on the front and back of the material, respectively. Otherwise, the insufficient force transmission may make the other side of the material receive less force, which will result in unsuccessful light detection. Furthermore, the device also requires a CCD for auxiliary recording, which needs an additional power supply and makes its overall volume to be relatively large. Some smaller devices for information writing and display, although simple to prepare, require additional high-energy devices such as ultrasound sources[10] for excitation. While others only need simple stretching or compression to realize the display of information, the luminescent material must be made into the shape of the information[11] during preparation and then cured into a transparent substrate, and only fixed information can be displayed during the display process. And it requires a complex fabrication process, and the shape needs to be changed in order to show the information[12]. Therefore, it will be better if the information storage and display device is passive, compact, simple and easy to fabricate..

How to achieve stronger luminescence of the mechanoluminescence material under low stress is another research focus. On the one hand, it can be achieved by reducing the luminescent material's luminescence threshold and increasing the efficiency of converting mechanical energy into light energy, such as developing new luminescence materials[13] [14]. On the other hand, stronger luminescence can be achieved through the auxiliary effect of the elastic substrate. For example, adding silica nanoparticles to the PDMS substrate increases the substrate's elastic modulus to improve the force transmission effect[15] and achieve stronger luminescence under the same deformation. However, in the existing research, little attention has been paid to use flexible materials with a small elastic modulus as the substrate, such as ecoflex. Here, we use ecoflex as the substrate and use the triboelectric effect to achieve strong luminescence under small force. It only needs less than 1N of force to swipe over the ecoflex cured with ZnS: Cu, and it can emit visible light under daylight.

Based on the above research considerations, a thin film that can store and display information and is very simple to prepare has been developed. The film can use the electrification effect to enhance mechanoluminescence significantly and can display and record the writing traces in real-time. After writing, it is in an invisible state, and then it can ultimately show the previous friction area under the excitation of water mist. We revealed the relationship between the force and speed of the friction process and the light intensity, discovered the relationship between the light intensity excited by the droplet and the polarity of the liquid, and the decay period of the triboelectric signal, and predicted the signal's retention time. Finally, we proposed the luminescence mechanism of the triboelectric effect enhanced mechanoluminescence and the luminescence mechanism of ML materials excited by droplets and extended it to the field of material contact luminescence.

## 2. Results and discussion

Figure 1a shows the preparation process of the functional film. It is straightforward. Zn: Cu powder and ecoflex are uniformly mixed and then cure, then a functional film capable of information storage and the display is obtained. Figure 1b first shows the information storage process. The information can be stored by sliding a glass rod or other friction pair on the film's surface. When the glass rod swipe on the surface of the functional film, bright green light emitted in the contact area, the emission wavelength is about 517nm, and the area after rubbing will continue to emit light for a period, which is related to the glow effect of ZnS: Cu and also to the residual charge in the friction area. After a time interval of about 1 second, all the light on the film will disappear, and the information written just now is invisible. When performing the information display process, spray water mist on the film, and traces of friction just now will be fully displayed, and its luminous intensity is related to the stroke strength and signing speed. After the water mist is wiped dry, the functional film is refreshed and reset, and a new round of writing and displaying operations can be performed.

In Figure 1b, it has been shown that only the rubbed area on the film will glow when the water droplets drop. Now the remaining questions are as follows. The first question is whether any liquid drops on the rubbed area can produce luminescence. The second question is whether there is a specific relationship between the luminous intensity and the liquid type. The third question is whether increasing the pressure, friction, and friction cycles can increase the luminous intensity excited when the liquid drops. To answer the above questions, we designed and built the experimental device and experimental process shown in Figure 2a, which can accurately control the friction process, the droplet size and drop location, and detect the light intensity excited by the droplet. First, we use a vertical electric stage to control the glass rod's depth into the sample and use a two-dimensional cantilever sensor to detect the pressure and friction in the vertical and horizontal directions in real-time and the two-dimensional displacement in the horizontal direction. The station controls the position of the functional film. Accurately control the penetration depth, pressure, friction force, and friction times of the glass rod rubbing sample at position 1. After the friction at position 1, the sample is transferred to position 2 through the stage's program control. While the sample is moving, the infusion needle installed in the springe pump drops the liquid accurately and uniformly on the friction surface at a speed of 0.5ml/min.

Figure 2b shows the relationship between excitation light intensity and liquid types. Six polar liquids are used, namely water, anhydrous alcohol, acetone, glycerin, acetic acid, and (OMIm), and four non-polar liquids, namely PAO 6, Liquid paraffin, isopropyl myristate, petroleum ether. It is found that only polar liquids can excite light, and the light intensity excited by the liquid dropping on the friction surface is related to the liquid's polarity—the greater the polarity, the stronger the light intensity that can be excited. Among them, the intensity of light that can be excited by water is the largest, and the intensity of light that can be excited by the weakest OmIm is the smallest.

Figure 2c shows the relationship between the excitation light intensity and the pressure and friction when the glass rod's sliding speed on the film surface is controlled at 23.7mm/s. It is found that the excitation light intensity of the liquid will increase with the increase in pressure. Moreover, there is probably a proportional relationship. It is not difficult to understand because when the pressure increases, the contact area between the glass rod and the film will increase, and the bonding distance at the microscopic scale will become smaller. According to the triboelectric transfer theory[16-18] proposed by Wang et al., the number of electrons transferred will increase significantly, so when the glass rod is scratched across the film's surface with more significant pressure and friction, more charges will remain on the surface. Considering the relationship between the liquid excitation light intensity and the polarity of the liquid and the frictional force's size, combined with the property of ZnS: Cu material, which shows that the stronger the field, the stronger the luminescence. The most reasonable explanation for droplets can excite the friction area to emit light is that when the liquid drops, an electric field is generated between the liquid drop and the friction surface to make the luminescent particles produce electroluminescence.

Figure 2d shows the relationship between the intensity of light that can be excited by the liquid and the number of frictions. The friction distance is 40mm each time, the pressing depth is 1mm, and the pressure is 1.2N±0.1N. It can be seen that the light intensity will increase with the increase in the number of frictions, but will decrease slightly after 4 to 5 times. This weak drop can be explained by the capacitor model. The process of glass rod and friction can be regarded as the charging process of the upper and lower plates of the capacitor. With the increase of the number of frictions, the charge on the upper and lower plates becomes saturated. When the two surfaces of dissimilar charges are in contact with each other, a discharge will consume a part of the charge, but the consumed charge will recover with the increase in the number of frictions, and then discharge, recover, and so on, which will also cause light intensity fluctuation.

Figure 3a shows the electrical potential signal distribution diagram after the film's surface is lightly scratched by the glass rod. It can be seen that there are prominent electrical potential peaks at the friction marks. Figure 3b is the distribution diagram of the signal's electric potential over time by continually sweeping the electrometer probe back and forth when the ambient humidity is 30%. The electric potential replication gradually decays with the increase of time, which can be well fitted by the exponential attenuation function. If it is considered that the signal stored in the film disappears when it decays to 5% of the initial potential signal, then it can be roughly inferred that the storage time of the written signal on the functional film in the 2000s, which is three times the fitted time constant. To extend the time of information storage, adjust the functional film storage environment's humidity. A more straightforward method is to put it in a vacuum for storage, lowering the potential dissipation speed by reducing gas molecules and water vapor molecules' influence to write signals, which greatly increases the signal storage time.

Figure 3c and Figure 3d show the factors affecting the luminous intensity of the glass rod sliding on the surface of the film. It exhibits that increasing the sliding speed and pressure can increase luminous intensity. But judging from the ratio of increasing the light intensity, increasing the sliding speed has a more obvious effect on improving the light intensity.

The spectral curves of different excitation modes are shown in Figure 4a, and it is also apparent that the positions of the peaks generated by different excitation modes are slightly shifted. This peak shift has been reported in many studies. The reason is that the transition metal Cu ion lacks 5s and 5p energy level electrons to shield the external field, so its excitation energy level 4f is environmentally sensitive and susceptible to host crystal[19] and the environment, resulting in a slight shift in the wave function, causing a shift in the center wavelength of the emission peak. Jeong et al. found that when the cured ZnS: Cu PDMS film is stretched and relaxed, the emission wavelength shift can reach 3nm under the effect of low and high operating rates[5]. In Figure 4a, the measured electroluminescence peak consists of two small peaks, 509nm and 455nm, corresponding to blue and green light components[5], which have also been reported in previous studies.

Figure 4b shows the mechanism of force luminescence enhanced by triboluminescence. When there is only force, the piezoelectric crystal ZnS: Cu[20] generates a local electric field. The conduction band and the valence band tilt in the electric field. Under the action of thermodynamics, the electrons in the defect level are more likely to escape to the conduction band, thereby recombining with the hole and emitting green light. since the different ability of the two materials to bind electrons, it is easy to transfer electrons when the glass rod slides on the functional film. Two sufaces with heterogeneous charges provides ZnS: Cu an additional electric field besides the local electric field generated by pressure, making the deflection of the conduction band and the valence band more prominent, more electrons will escape to the conduction band due to thermodynamics, more electrons and holes will recombine, therefore this tribological process can produce stronger light. This enhanced electrification phenomenon is particularly evident in Figure 5b. When the glass rod slides on the film at a low speed of 1.8mm/s, the light intensity is only about 1/10 of the glass rod back against the surface. The sliding speed of writing is not equivalent to the luminous intensity until the speed increases to 23.7mm/s.

Figure 4d is the mechanism diagram of the droplet excitation. When the polar molecules gradually approach the rubbed area, they will transit from disorder to order. An electric field will then be generated between the rubbed surface and the polar liquid, thereby ZnS: Cu particles on the film surface will emit light in the electric field.

Similarly, this observation has also been found to apply to other solids. For example, when the glass rod is lifted on the surface of the film, the light intensity is much greater than that of the sliding process, which demonstrats the electroluminescence effect enhanced by triboelectricity. The luminescence mechanism is similar to that of droplet luminescence, and the field strength between two surfaces with different kinds of charges excites ZnS: Cu particles to produce luminescence.

## 3. Conclusion

In summary, a flexible, functional film that can be used for information writing and display is prepared out. It uses ecoflex with a very small elastic modulus as the matrix of ZnS: Cu particles, and uses the triboelectric effect to achieve enhanced mechanoluminescence. Stronger luminescence of ZnS: Cu under the same force is reached. This device uses liquid droplets to excite luminescence and shows triboelectricity, electroluminescence, and electroluminescence. Its memory effect can record the traces left by the friction pair on it in real-time through the triboelectric effect, and then the friction area can also be displayed completely when the polar liquid is falling. We explored its influencing factors and proposed a mechanism explanation. The device is expected to play a role in display, information storage, and encryption.

## 4. Experimental Section

Material: *The doped ZnS material was purchased from Shanghai Keyan Instrument Co., Ltd. The composition and crystal state were characterized by X-ray diffraction (XRD, D8 Advance, Bruker) and XPS (PHI Quantera II, Ulvac-Phi Inc.), revealing that the surface of the ZnS: Cu particles were evenly covered with a layer of Al2O3. The XRD results show that it is a better zink blende crystal phase.*

Preparation of functional film: *The doped ZnS material and ecoflex 30 are uniformly mixed at a mass ratio of 1:3 and then introduced into a glass petri dish, and then placed in a drying oven for 30 minutes.*

Optical characterization: *A photon counter (C8855-01, Hamamatsu) is used to characterize the light intensity quantitatively. The 0.1nm resolution spectrometer (HRS-300SS) is used to detect the spectrum, and an optical fiber with a 1mm diameter receiving range is used to transmit light. The experiment video was taken by an electronic video camera (EOS 5D Mark III), and the experiment pictures were captured from the video.*

Light intensity measurement of liquid drip-induced luminescence: *The device consists of a three-dimensional electric stage (Daheng Optoelectronics Co., Ltd.), a glass rod sliding friction pair fixed in the Z-axis direction, a single photon counter, and a springe Pump (chemyx Inc). One side of the device can precisely control the friction process, such as friction force, friction position, and friction times. The other can accurately measure the light intensity excited by the liquid drop. One side of the optical fiber is used to transmit the light excited by the liquid drop precisely aims at the liquid's drop location and the other side is connected to a single photon counter (C8855-01, Hamamatsu) to record the light intensity. The medical infusion needle is aligned, and the springe pump program controls the droplet size. In this experiment, the droplet dropping speed is 0.5ml/min, and the exposure time for each data point of the light intensity is 10ms.*

Potential measurement: *The electric potential distribution surface scanning after the light-emitting film is drawn obtained by a combination of an electrometer (6517B Electrometer, Keithley) and a two-dimensional electronically controlled displacement platform with a control accuracy of 1um (Daheng Optoelectronics Co., Ltd.).*

Force measurement: *The force in the friction process is completed by combining the supporting device and the bracket, the electronically controlled translation stage, the cantilever sensor, and the transmitter, the NI acquisition card, and Labview software.*

**Acknowledgements**

This research was supported by the National Natural Science Foundation of China (51922058, 51675297, 51527901).

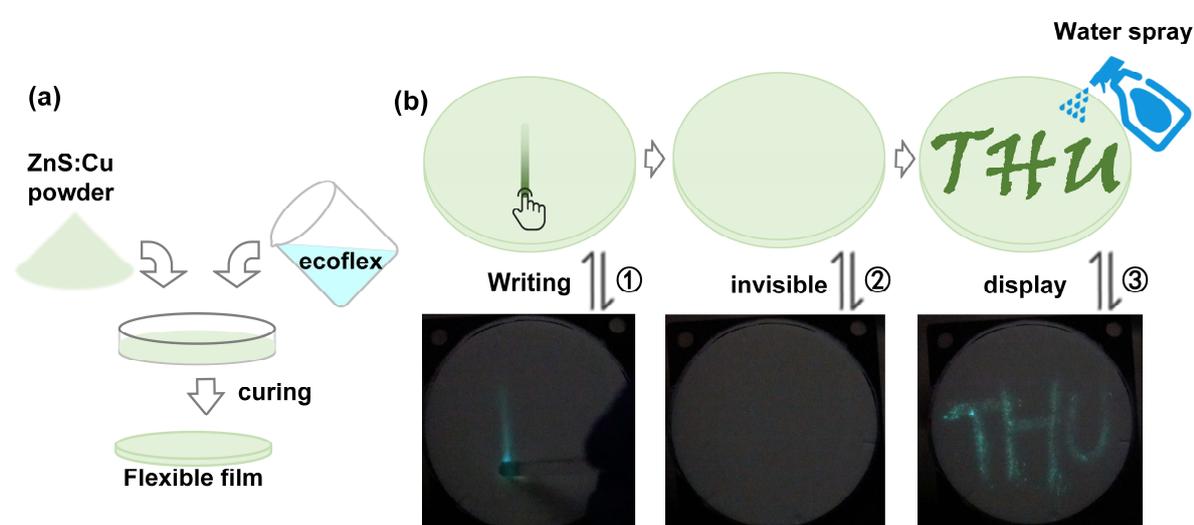

**Figure 1** The preparation and demonstration effect of the information writing device and the corresponding picture of the actual object a) The preparation process of the functional film b) From left to right are 1) the demonstration diagram and the actual operation diagram of the information writing process, holding the glass rod on the film by hand Write the information need to keep 2) The written information is in an invisible state 3) Memory effect: when the sprayed water drops drop onto the surface of the functional film with hidden information, the information is displayed entirely.

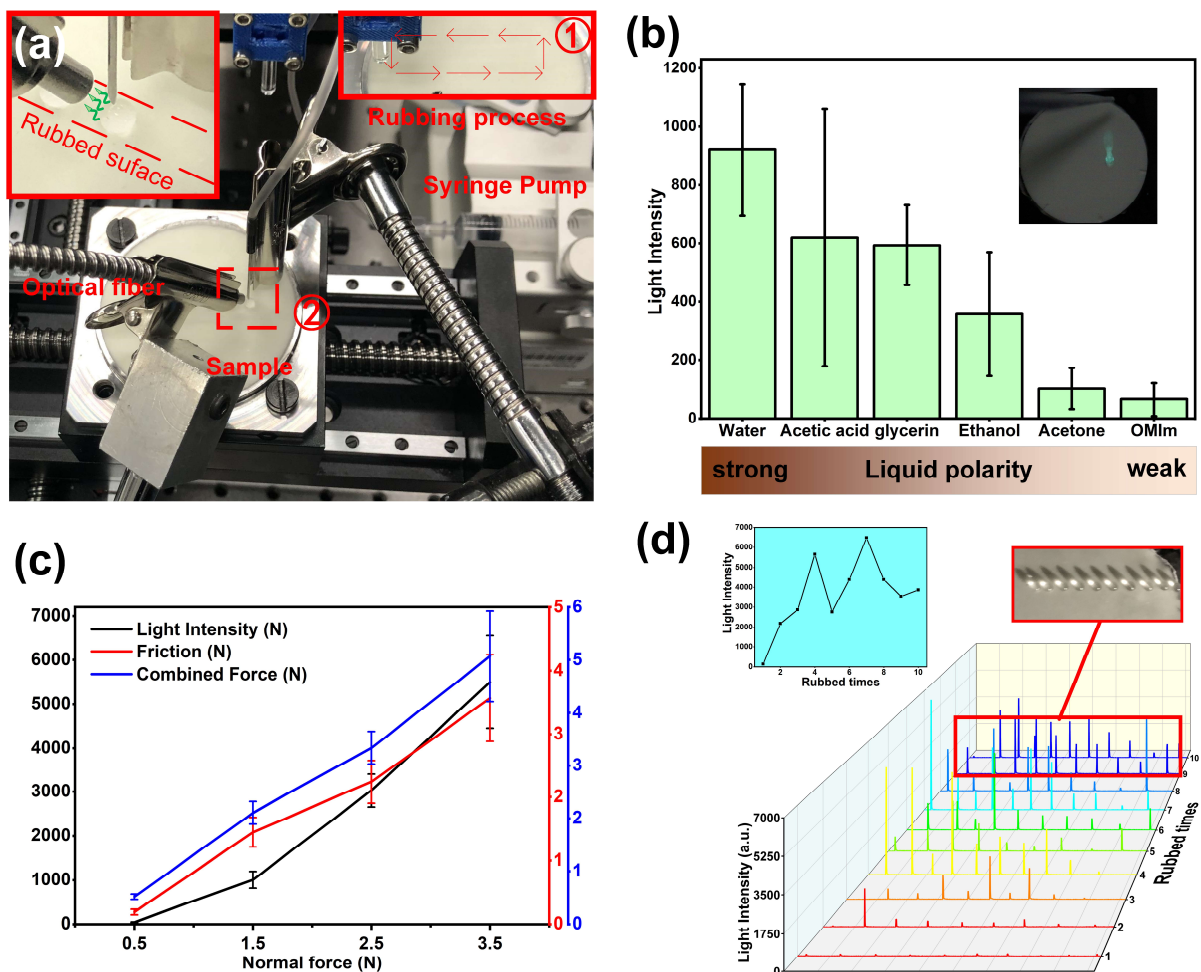

**Figure 2** Factors affecting the intensity of the excitation light when the droplet falls. a) An experimental device for quantitatively measuring the luminescence intensity excited by the droplet, consisting of three parts. The first part is composed of a 3D electronically controlled translation stage, holder, and sensor. The glass rod is fixed on the Z-axis electronically controlled stage by a particular clamping device, and it is parallel to the Z-axis. When the X-axis and Y-axis electronically controlled stage drive the film to move under the glass rod, a friction process occur. During the friction process, the friction force is recorded in real-time by a two-dimensional cantilever beam sensor at the end of the glass rod holder. The second part is composed of a medical infusion needle tube, infusion tube holder, and springe pump. The medical infusion tube is fixed on the holder, and the droplet size is controlled by the springe pump. The third part is composed of fiber, fiber holder, and photon counter. The optical fiber is fixed on the optical fiber holder to precisely align the droplet's landing point, collect the droplet position's luminescence, and then transmit it to the photon counter system to quantitatively characterize the intensity of the light excited by the droplet. b) The relationship between the light intensity generated when the liquid drops and the type of liquid. The liquid's polarity decreases from left to right, and the light intensity that the liquid can excite is positively correlated with the liquid's polarity. c) The light intensity is generated when the liquid drops are nearly proportional to the normal force between the glass rod and the functional film. d) The relationship between the light intensity excited by the droplet and the friction times between the glass rod and the functional film, where the positive pressure is 0.5N, the sliding speed is 23.7mm/s, and the sliding distance is 40mm. The small inset on the left is the right view of the three-dimensional image. The light intensity excited by the droplet fluctuates with the number of frictions. The small illustration on the right is a physical display of the water droplets falling

on the glass rod's rubbing marks. The number of water droplets is the number of peaks detected in the experimental light intensity curve.

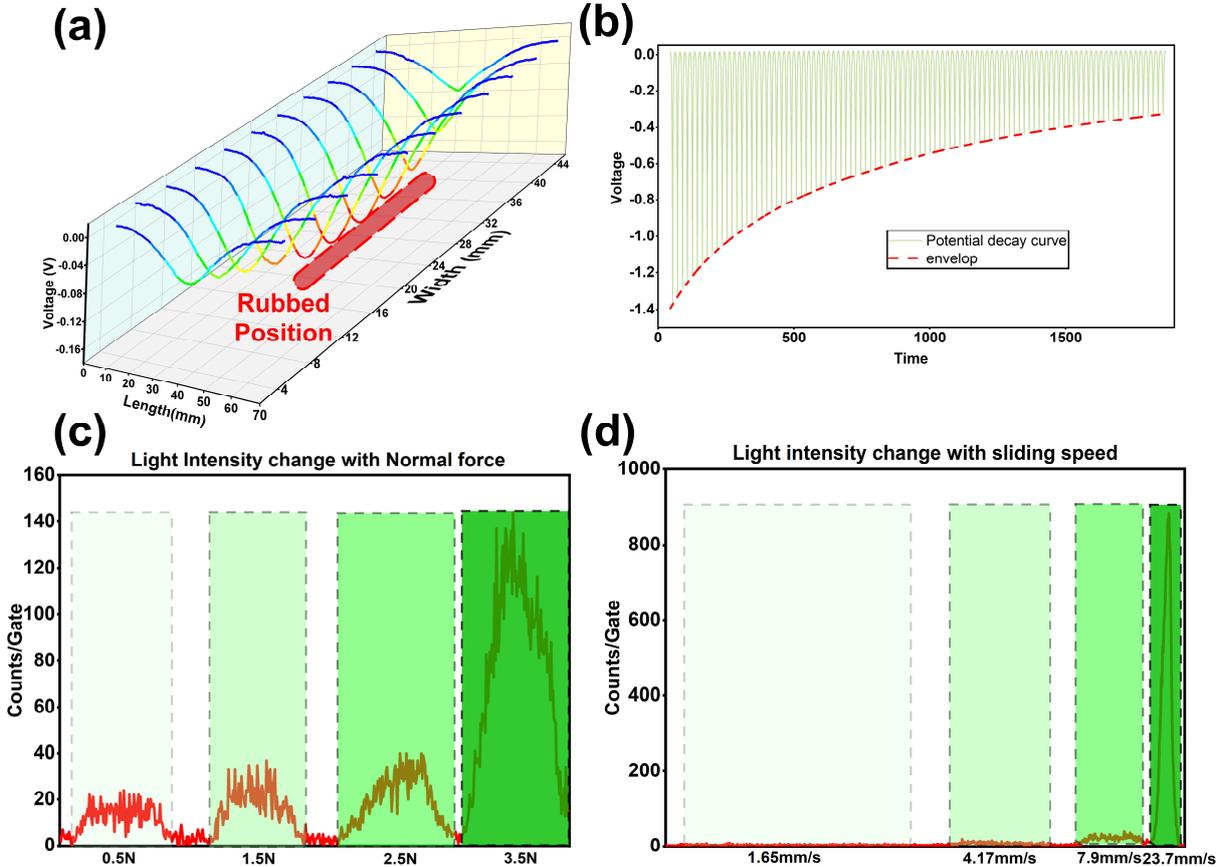

**Figure 3** The glass rod's influence on the surface potential after sliding on the functional film and the influencing factors of the light intensity excited by the glass rod. a) Use a glass rod with a diameter of 4mm to lightly traverse a 20mm straight line on the functional film surface, and the probe of the potentiometer sweeps a potential curve on the surface of the sample every 4mm to form a two-dimensional potential distribution diagram on the film surface. b) The attenuation curve of the electrical signal left by the glass rod across the functional film. c) The relationship between the light emitted by the glass rod sliding on the functional film and the normal force. d) The relationship between the light emitted by the glass rod sliding on the functional film and the sliding speed.

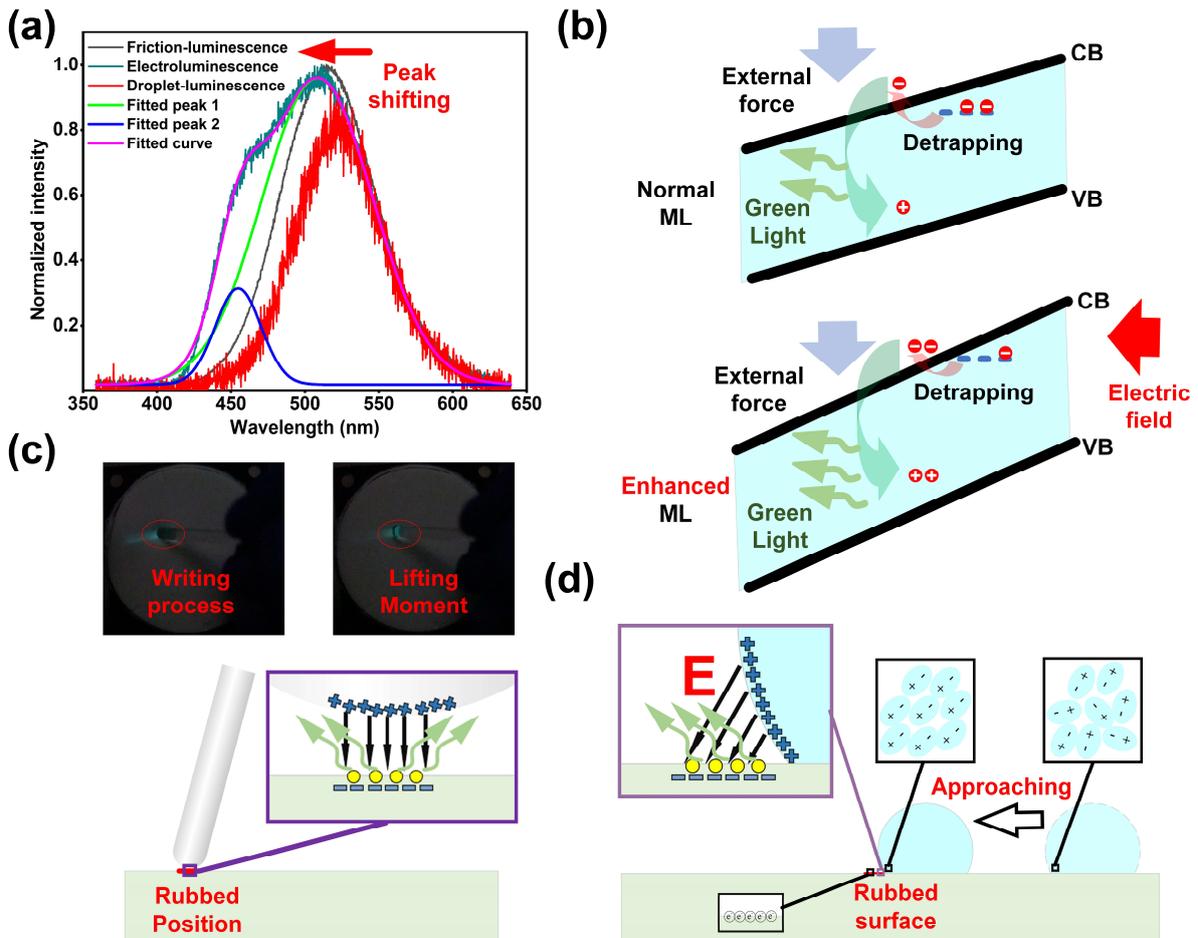

**Figure 4** Explanation of the mechanism of the surface luminescence of the functional film excited by the droplet

a) The positions of the luminescence peaks of different excitation processes. The electroluminescence peak is composed of two peaks with center wavelengths of 455nm and 509nm. The positions of the luminescence peaks produced by different excitation methods will be slightly shifted. b) Diagram of the explanation mechanism of triboelectric enhanced electroluminescence. The above is the pure mechanoluminescence diagram. The following is a diagram of the mechanism of enhanced triboelectric luminescence. c) The rubbed solids can also emit light when they touch or leave the rubbing area. The illustrations show the moments of writing and lifting the glass rod. It can be seen that at the moment of lifting, the light intensity of the excitation light is several times that of the glass rod when writing. d) An explanation diagram of the mechanism by which the droplets excite the friction area to produce light. The illustration shows the orientation change of the polar liquid from right to left.

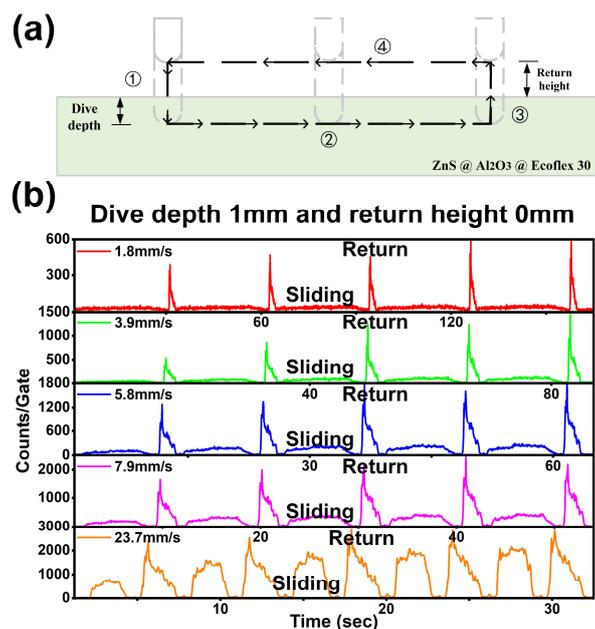

**Figure 5** The effect of electrification enhancement during the friction process of the glass rod is shown. a) The trajectory of the glass rod on the surface of the functional film b) The glass rod traverses the surface of the functional film at different speeds at a depth of 1 mm, and then returns along the original path of the friction surface, repeating the cycle five times. The figure shows a schematic diagram of the light intensity with the friction process with an integration time of 50ms.